\documentclass[draft,showpacs,preprintnumbers,amsmath,amssymb]{revtex4}

\begin{document}
\title{Topological Constraints at the Theta Point:  Closed Loops at Two Loops}
\author{William Kung}
\affiliation{Department of Physics and Astronomy, University of
Pennsylvania, Philadelphia, PA 19104-6396}
\author{Randall D. Kamien}\email{kamien@physics.upenn.edu}
\affiliation{Department of Physics and Astronomy, University of
Pennsylvania, Philadelphia, PA 19104-6396}

\date{\today}
\begin{abstract}

We map the problem of self-avoiding random walks in a $\Theta$ solvent
with a chemical potential for writhe to the three-dimensional
symmetric $U(N)$-Chern-Simons theory as $N\rightarrow 0$. We find a new scaling regime of topologically constrained polymers, with
critical exponents that depend on the chemical potential for writhe, which gives way to a fluctuation-induced first-order transition.
\end{abstract}

\maketitle

%
%
%
%
\input{epsf.sty}
%
%
%
%
%
%
%
%
%
%
%
%
%
%
%
%
%
%
%
%
%
%
%
%
%
%
%
%
%
%
%
%
%
%
%
%
%
%
%
%
%
%
%
%
%
%
%
%
%
%
%
%
%
%
%
%
%
%
%
%
%
%
%
%
%
%
%
%
%
%
%
%
%

%
%
%
%

%

%
%
%
%
%
%
\def\axowidth{0.5 }
\def\axoscale{1.0 }
\def\axoxoff{0 }
\def\axoyoff{0 }
\def\axoxo{0 }
\def\axoyo{0 }
\def\firstcall{1}
\def\Gluon(#1,#2)(#3,#4)#5#6{
%
%
\put(\axoxoff,\axoyoff){
}
}
\def\Photon(#1,#2)(#3,#4)#5#6{
%
%
\put(\axoxoff,\axoyoff){
}
}
\def\ZigZag(#1,#2)(#3,#4)#5#6{
%
%
\put(\axoxoff,\axoyoff){
}
}
\def\PhotonArc(#1,#2)(#3,#4,#5)#6#7{
%
%
\put(\axoxoff,\axoyoff){
}
}
\def\GlueArc(#1,#2)(#3,#4,#5)#6#7{
%
%
\put(\axoxoff,\axoyoff){
}
}
\def\ArrowArc(#1,#2)(#3,#4,#5){
%
%
\put(\axoxoff,\axoyoff){
}
}
\def\LongArrowArc(#1,#2)(#3,#4,#5){
%
%
\put(\axoxoff,\axoyoff){
}
}
\def\DashArrowArc(#1,#2)(#3,#4,#5)#6{
%
%
\put(\axoxoff,\axoyoff){
}
}
\def\ArrowArcn(#1,#2)(#3,#4,#5){
%
%
\put(\axoxoff,\axoyoff){
}
}
\def\LongArrowArcn(#1,#2)(#3,#4,#5){
%
%
\put(\axoxoff,\axoyoff){
}
}
\def\DashArrowArcn(#1,#2)(#3,#4,#5)#6{
%
%
\put(\axoxoff,\axoyoff){
}
}
\def\ArrowLine(#1,#2)(#3,#4){
%
%
\put(\axoxoff,\axoyoff){
}
}
\def\LongArrow(#1,#2)(#3,#4){
%
%
\put(\axoxoff,\axoyoff){
}
}
\def\DashArrowLine(#1,#2)(#3,#4)#5{
%
%
\put(\axoxoff,\axoyoff){
}
}
\def\Line(#1,#2)(#3,#4){
%
%
\put(\axoxoff,\axoyoff){
}
}
\def\DashLine(#1,#2)(#3,#4)#5{
%
%
\put(\axoxoff,\axoyoff){
}
}
\def\CArc(#1,#2)(#3,#4,#5){
%
%
\put(\axoxoff,\axoyoff){
}
}
\def\DashCArc(#1,#2)(#3,#4,#5)#6{
%
%
\put(\axoxoff,\axoyoff){
}
}
\def\Vertex(#1,#2)#3{
%
%
\put(\axoxoff,\axoyoff){
}
}
\def\Text(#1,#2)[#3]#4{
%
%
\dimen0=\axoxoff \unitlength \dimen1=\axoyoff \unitlength
\advance\dimen0 by #1 \unitlength \advance\dimen1 by #2
\unitlength \makeatletter \@killglue\raise\dimen1\hbox
to\z@{\kern\dimen0 \makebox(0,0)[#3]{#4}\hss} \ignorespaces
\makeatother }
\def\BCirc(#1,#2)#3{
%
%
\put(\axoxoff,\axoyoff){
}
}
\def\GCirc(#1,#2)#3#4{
%
%
\put(\axoxoff,\axoyoff){
}
}
\def\EBox(#1,#2)(#3,#4){
%
%
\put(\axoxoff,\axoyoff){
}
}
\def\BBox(#1,#2)(#3,#4){
%
%
\put(\axoxoff,\axoyoff){
}
}
\def\GBox(#1,#2)(#3,#4)#5{
%
%
\put(\axoxoff,\axoyoff){
}
}
\def\Boxc(#1,#2)(#3,#4){
%
%
\put(\axoxoff,\axoyoff){
}
}
\def\BBoxc(#1,#2)(#3,#4){
%
%
\put(\axoxoff,\axoyoff){
}
}
\def\GBoxc(#1,#2)(#3,#4)#5{
%
%
\put(\axoxoff,\axoyoff){
}
}
\def\SetWidth#1{\def\axowidth{#1 }}
\def\SetScale#1{\def\axoscale{#1 }}
\def\SetOffset(#1,#2){\def\axoxoff{#1 } \def\axoyoff{#2 }}
\def\SetScaledOffset(#1,#2){\def\axoxo{#1 } \def\axoyo{#2 }}
\def\pfont{Times-Roman }
\def\fsize{10 }
\def\SetPFont#1#2{\def\pfont{#1 } \def\fsize{#2 }}
%
%
\makeatletter
\def\fmode{4 }
\def\@l@{l} \def\@r@{r} \def\@t@{t} \def\@b@{b}
\def\mymodetest#1{\ifx#1\end \let\next=\relax \else {
\if#1\@r@\global\def\fmodeh{-3 }\fi \if#1\@l@\global\def\fmodeh{3
}\fi \if#1\@b@\global\def\fmodev{-1 }\fi
\if#1\@t@\global\def\fmodev{1 }\fi } \let\next=\mymodetest\fi
\next} \makeatother
\def\PText(#1,#2)(#3)[#4]#5{
%
%
\def\fmodev{0 }
\def\fmodeh{0 }
\mymodetest#4\end
\put(\axoxoff,\axoyoff){\makebox(0,0)[]{\special{"/\pfont findfont
\fsize
 scalefont setfont #1 \axoxo add #2 \axoyo add #3
\fmode \fmodev add \fmodeh add \fsize (#5) \axoscale ptext }}} }
\def\GOval(#1,#2)(#3,#4)(#5)#6{
%
%
\put(\axoxoff,\axoyoff){
} }
\def\Oval(#1,#2)(#3,#4)(#5){
%
%
\put(\axoxoff,\axoyoff){
} }
\let\eind=]
\def\DashCurve#1#2{\put(\axoxoff,\axoyoff){
}}
\def\Curve#1{\put(\axoxoff,\axoyoff){
}}
\def\kromme(#1,#2)#3{#1 \axoxo add #2 \axoyo add \ifx #3\eind\else
\expandafter\kromme\fi#3}
\def\LogAxis(#1,#2)(#3,#4)(#5,#6,#7,#8){
%
%
\put(\axoxoff,\axoyoff){
}
}
\def\LinAxis(#1,#2)(#3,#4)(#5,#6,#7,#8,#9){
%
%
\put(\axoxoff,\axoyoff){
}
}
\input rotate.tex
\makeatletter
\def\rText(#1,#2)[#3][#4]#5{
%
%
\ifnum\firstcall=1\global\def\firstcall{0}\rText(-10000,#2)[#3][]{#5}\fi
\dimen2=\axoxoff \unitlength \dimen3=\axoyoff \unitlength
\advance\dimen2 by #1 \unitlength \advance\dimen3 by #2
\unitlength \@killglue\raise\dimen3\hbox to \z@{\kern\dimen2
\makebox(0,0)[#3]{ \ifx#4l{\setbox3=\hbox{#5}\rotl{3}}\else{
\ifx#4r{\setbox3=\hbox{#5}\rotr{3}}\else{
\ifx#4u{\setbox3=\hbox{#5}\rotu{3}}\else{#5}\fi}\fi}\fi}\hss}
\ignorespaces } \makeatother
\def\BText(#1,#2)#3{
%
%
\put(\axoxoff,\axoyoff){
}
}
\def\GText(#1,#2)#3#4{
%
%
\put(\axoxoff,\axoyoff){
}
}
\def\B2Text(#1,#2)#3#4{
%
%
\put(\axoxoff,\axoyoff){
}
}
\def\G2Text(#1,#2)#3#4#5{
%
%
\put(\axoxoff,\axoyoff){
}
}

\def\Lk{\hbox{\sl Lk}}
\def\Wr{\hbox{\sl Wr}}
\def\Tw{\hbox{\sl Tw}}

The statistics and properties of random walks are central to our understanding of
a large class of phenomena in mathematics, biology, physics and even economics.  The broad applicability of random walks in modelling physical situations
 is, in part, due to the Gaussian
statistics of its correlations, rendering it attractive from an
analytical viewpoint.  In polymer physics, self-avoiding random
walks model polymers in a good solvent that are much longer than
their persistence length. This identification establishes a theoretical
basis for the computation of scaling exponents which are, in turn,
related to the exponent $\nu$ that relates the radius of gyration
$R_G$ to the length of the chain $L$ through $R_G \sim L^{\nu}$.
In this letter we study the scaling behavior of closed polymers
with a chemical potential for their writhe \Wr.  Closed loops of
DNA, as found in bacterial plasmids, experience a topological
linking number constraint enforced by the twist rigidity of the
double backbone. Refinements can be made to de Gennes' mapping to
incorporate this additional physics through the addition of a
minimal coupling to a Chern-Simons gauge field \cite{Kamien97}.
Specifically, the $N\rightarrow 0$ limit of a $U(N)$ Chern-Simons
theory with the gauge field coupled to the common $U(1)$ can be
used to study self-avoiding walks with this topological
constraint. The Abelian Chern-Simons gauge field is used to
enforce the topological constraint in the form of White and
Fuller's \cite{White69, Fuller71} famous relation
\cite{Marko95,Fain97a,Fain97b,Julicher94} $\Lk=\Tw+\Wr$. It
relates the topological linking number $\Lk$ to the total amount
of twist $\Tw$ and writhe $\Wr$ found in, for example, DNA
backbones.  We find that the topological constraint leads to a new
universality class of random walks.

Without a topological constraint, the individual monomers suffer
two interactions: a chemical interaction which depends on the
solvent and a hard-core interaction which is entropic. In the case
of good solvents, these interactions of chain connectivity and
excluded volume are described by de Gennes' mapping
\cite{Gennes72} of the $O(N)$-symmetric $\phi^4$-theory in the
$N\rightarrow0$ limit. The Wilson-Fisher fixed point controls the
scaling in $d=4-\epsilon$ dimensions where $\nu\approx 0.588$ in $d=3$
\cite{Kamien97}. In poor solvents, the polymers collapse into
compact globules and $R_G \sim L^{\frac {1}{d}}$. The transition
between the two solvent regimes is the $\Theta$ point which is
described by an $O(N)$-symmetric $\phi^6$-model at the point where
the $\phi^4$ term vanishes or, in other words, at the tricritical
point \cite{Gennes75}. Logarithmic corrections to mean-field
results have been calculated, but it has been proven difficult to
test these predictions experimentally \cite{Duplantier82,
Duplantier87}.  Computer simulations \cite{Grassberger, Young,
Rubio} have also been performed but agreement with theory has only
been tentative.

The topological constraint imposed by \Lk\ is inherently
three-dimensional as is the Chern-Simons term which surrogates for
it.  As a result, the perturbative study of a polymer in a good
solvent cannot be performed via a conventional $\epsilon=4-d$
expansion. In \cite{Kamien97} it was shown that the exponents at
the $d=3$ Wilson-Fisher fixed point were unchanged at one loop
order. Because there is no controlled approximation to go to
higher order, we have focussed here on polymers at the $\Theta$
point with a topological constraint so that a systematic expansion
can be performed directly in $d=3-\epsilon$ dimensions.

At the $\Theta$ point, the exponent $\nu$ is the correlation length exponent at the tricritical point $\mu=u=0$ of the following free energy density for the $N$-component, complex scalar $\vec\phi$
\cite{Gennes72,Gennes75}:
\begin{equation}
f=\vert\partial_{\mu}\phi_i\vert^2+\mu\vert\vec\phi\vert^2+u\left(\vert\vec\phi\vert^2\right)^2
+w\left(\vert\vec\phi\vert^2\right)^3, \label{langragian}
\end{equation}
in the $N\rightarrow 0$ limit.  In
three dimensions, renormalization group analysis shows that
although there are logarithmic corrections to mean-field behavior,
the scaling exponents agree simply with predictions from
dimensional analysis \cite{Hager99, Hager02}.  To incorporate the
topological constraints found in closed polymers, we introduce an
Abelian Chern-Simons gauge field with the free energy density
\cite{Kamien97}:
\begin{equation}
f=\frac{1}{2}\,\epsilon^{\mu\nu\rho}A_{\mu}\partial_{\nu}A_{\rho}.
\label{ChernSimonL}
\end{equation}
Since we are interested in the critical behavior of our theory
near the $\Theta$ point, we perform our analysis at the
tricritical point where $\mu$ and $u$ vanish. The energy is the
sum $F=F_{\rm{CS}}+F_{\rm{b}} +F_{\rm{gf}}$ where
\begin{eqnarray}
F_{\rm{CS}}&=&\frac{1}{2}\int\,{\mathrm
d}^3x\,\epsilon_{\mu\nu\rho}A_{\mu}\partial_{\nu}A_{\rho}\\
F_{\rm{b}}&=&\int\,{\mathrm d}^3x \left\vert\partial_{\mu}
\phi_{i}-ig_0A_{\mu}\phi_{i}\right\vert^2+V(\phi_{i})\\
F_{\rm{gf}}&=&\frac{1}{2\Delta}\int{\mathrm d}^3x
(\partial_{\mu}A_{\mu})^2\\
V(\phi_{i})&=&\mu_0\vert\vec\phi\vert^2+w_0\left(\vert\vec\phi\vert^2\right)^3
\end{eqnarray}
We impose the Landau gauge $(\Delta\rightarrow 0)$ in
$F_{\rm{gf}}$ for all our subsequent calculations.  Because the graphs first
diverge at two-loops, standard dimensional regularization is adequate and
leads to $\frac{1}{\epsilon}$ poles in $d=3-\epsilon$ dimensions.
To evaluate each two-loop diagram, we perform the first momentum integral in $d=3$ and
then use dimensional regularization on the remaining single integral.  This scheme works because
the first integral can only give power law divergences since any integrand with odd powers of momentum vanishes, leaving only even powers of the momenta compared to the three-dimensional measure.   Moreover, by power-counting, the remaining integral must diverge logarithmically and cannot cancel the power-law divergence from the first integration.  Thus the logarithms only arise in the
second of the two integrations.
A more sophisticated
treatment suggests that this scheme is consistent to all orders \cite{silva00}.  As
is usual, we first perform the necessary tensor algebra in
physical dimensions before analytically continuing the dimensions
of the resulting scalar integrand.  The correlation functions
satisfy the Slavnov-Taylor identities
\cite{Chen92} and the Ward identities \cite{silva00};  gauge
invariance is therefore preserved.

\bigbreak
We first introduce the renormalized free energy as follows:
\begin{eqnarray}
F&=&\int\,{\mathrm
d}^3x\,\Big\{Z_{\phi}|\partial_{\mu}\phi_{i}|^2+\tilde \mu Z_\mu\vert\vec\phi\vert^2-i\tilde gZ'_g\,[(A_{\mu}\phi_{i})^{*}\partial_{\mu}\phi_{i} +(\partial_{\mu}\phi_{i}^{*})A_{\mu}\phi_{i}] + Z''_{g}\,\tilde g^2\left|A_{\mu}\phi\right|^2\nonumber\\
&&+\frac{1}{2}Z_{A}\,\epsilon^{\mu\nu\rho}
A_{\mu}\partial_{\nu}A_{\rho}+\frac{1}{2\Delta}(\partial_{\mu}A^{\mu})^{2}
+Z_w\tilde w\left(\vert\vec\phi\vert^2\right)^3\Big\},
\end{eqnarray}
where $\tilde g = g M^{(3-d)/2}$, $\tilde \mu = \mu M^2$,  $\tilde w = w M^{6-2d}$, and
$M$ is the momentum scale at which we are renormalizing.
Denoting $\Gamma_R^{(N_\phi,N_A)}$ as the renormalized proper vertex of $N_\phi$
scalar fields and $N_A$ gauge fields, the Callan-Symanzik equation
is:
\begin{eqnarray}
\bigg[M\frac{\partial}{\partial
M}+\beta_{\mu}\frac{\partial}{\partial\mu
}+\beta_{w}\frac{\partial}{\partial
w}+\beta_{g}\,\frac{\partial}{\partial
g}-\frac{1}{2}N_\phi\eta_{\phi}
-\frac{1}{2}N_A\eta_{A}\bigg]\,\Gamma_R^{(N_\phi,N_A)}=0,
\end{eqnarray}
where $M$ is the renormalization scale.  The engineering dimension
of any correlation function $\Gamma$ depends on its field content
so that $d(\Gamma)=3-\frac{1}{2}N_{\phi}-N_A$.  We have
\begin{eqnarray}
\eta_\phi &=& M\frac{\partial}{\partial M} \ln Z_\phi\\
\eta_A &=& M\frac{\partial}{\partial M} \ln Z_A\\
\mu_0 &=& \mu M^2 \frac{Z_\mu}{Z_\phi}\\
g_0 &=&  g M^{(3-d)/2} \frac{Z'_g}{Z_\phi\sqrt{Z_A}}\\
w_0&=& wM^{6-2d}\frac{Z_w}{Z^3_\phi}
\end{eqnarray}
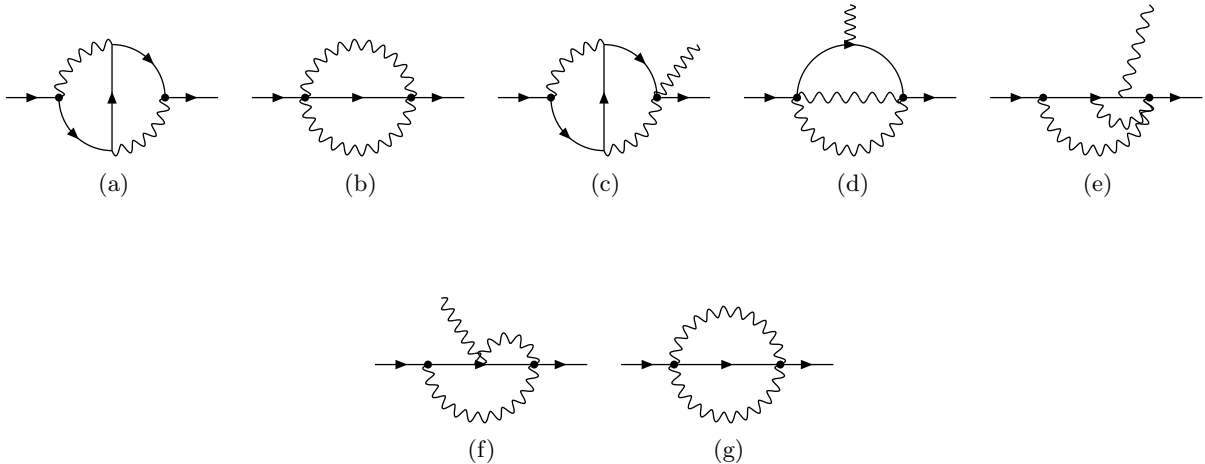
\begin{figure}
\begin{picture}(90,100)(0,0)
\PhotonArc(50,50)(20,90,180){2}{6}
\PhotonArc(50,50)(20,270,360){2}{6} \Vertex(30,50){1.5}
\Vertex(70,50){1.5} \ArrowLine(10,50)(30,50)
\ArrowLine(70,50)(90,50) \ArrowLine(50,30)(50,70)
\ArrowArc(50,50)(20,180,270) \ArrowArc(50,50)(-20,180,270)
\put(45,15){(a)}
\end{picture}
\begin{picture}(90,100)(0,0)
\ArrowLine(10,50)(30,50) \ArrowLine(30,50)(70,50)
\ArrowLine(70,50)(90,50) \PhotonArc(50,50)(20,180,0){2}{12}
\PhotonArc(50,50)(20,360,180){2}{12} \Vertex(30,50){1.5}
\Vertex(70,50){1.5} \put(45,15){(b)}
\end{picture}
\begin{picture}(90,100)(0,0)
\PhotonArc(50,50)(20,90,180){2}{6}
\PhotonArc(50,50)(20,270,360){2}{6} \Vertex(30,50){1.5}
\Vertex(70,50){1.5} \ArrowLine(10,50)(30,50)
\ArrowLine(70,50)(90,50) \ArrowLine(50,30)(50,70)
\ArrowArc(50,50)(-20,180,270) \ArrowArc(50,50)(20,180,270)
\Photon(70,50)(85,70){2}{6} \put(45,15){(c)}
\end{picture}
\begin{picture}(90,100)(0,0)
\ArrowLine(10,50)(30,50) \Photon(30,50)(70,50){2}{6}
\Photon(50,70)(50,85){2}{4} \ArrowLine(70,50)(90,50)
\ArrowArcn(50,50)(20,180,0) \PhotonArc(50,50)(20,180,360){2}{12}
\Vertex(30,50){1.5} \Vertex(70,50){1.5} \put(45,15){(d)}
\end{picture}
\begin{picture}(90,100)(0,0)
\ArrowLine(10,50)(30,50) \ArrowLine(30,50)(70,50)
\ArrowLine(70,50)(90,50) \PhotonArc(60,50)(10,180,360){2}{6}
\PhotonArc(50,50)(20,180,360){2}{12} \Photon(60,50)(70,85){2}{6}
\Vertex(30,50){1.5} \Vertex(70,50){1.5} \put(45,15){(e)}
\end{picture}
\begin{picture}(90,100)(0,0)
\ArrowLine(10,50)(30,50) \ArrowLine(30,50)(70,50)
\ArrowLine(70,50)(90,50) \PhotonArc(60,50)(10,0,180){2}{6}
\PhotonArc(50,50)(20,180,360){2}{12} \Photon(50,50)(35,75){2}{6}
\Vertex(30,50){1.5} \Vertex(70,50){1.5} \put(45,15){(f)}
\end{picture}
\begin{picture}(90,100)(0,0)
\ArrowLine(10,50)(30,50) \ArrowLine(30,50)(70,50)
\ArrowLine(70,50)(90,50) \PhotonArc(50,50)(20,180,0){2}{12}
\PhotonArc(50,50)(20,360,180){2}{12} \Vertex(30,50){1.5}
\Vertex(70,50){1.5} \put(45,15){(g)}
\end{picture}
\caption{Two-loop diagrams arising from the gauge field coupling
to the scalar field:  contributions to $Z_{\phi}$ (a),(b);
corrections to the cubic gauge vertex (c)-(f); contributions to
$Z_{\mu}$(p=0) (g).} \label{2LoopMatter}
\end{figure}

Perturbatively, the underlying symmetry of our action provides a
tight constraint on the divergences in our calculations.  The odd
parity of the Chern-Simons field prevents any correction to the
scalar field at the one-loop level: by rescaling $A_\mu \rightarrow g^{-1} A_\mu$, we see that under
parity $g^2\rightarrow -g^2$, and so correlations of $\phi$ which are parity invariant can only depend on $g^4$ and thus the first corrections are at two loops.
Accordingly, we do not find any one-loop contributions to $Z_\phi$ or $Z_\mu$.
Further, the Coleman-Hill
theorem shows that the $\beta$-function of the Chern-Simons gauge
coupling receives no contribution beyond one-loop
\cite{Semenoff88,Niemi83,Redlich84,Coleman85} in perturbation.
More simplification results from the vanishing of diagrams at
any order with closed scalar loops:  they necessarily introduce a
combinatoric factor of $N$ and do not provide corrections to the
gauge field as $N\rightarrow 0$ and so a Maxwell term,
$F_{\mu\nu}^2$, though allowed by symmetry, is not generated.
The details of the combinatoric argument go as
follows: consider a graph with only external gauge-field legs.
Since an external gauge-field leg must necessarily connect to two
internal $\phi$ legs (and possibly one internal gauge-field leg)
and since the $U(N)$ index of $\phi$ is not carried by any of the
other external legs, there must be a sum over that index. Since
that sum is proportional to $N$, this graph vanishes as
$N\rightarrow 0$. More complex internal topologies only {\sl add}
factors of $N$ to the graph and do not change this result,
thus $Z_A=1$. For the same reason there exists only a handful of
potential two-loop contributions to the renormalization functions
$Z_X$.

All the non-vanishing two-loop Feynman diagrams are shown in
Figures 1 and 2.  Figs. 1a and 1b are contributions to $Z_{\phi}$,
while Fig. 1c through 1f are corrections to the cubic gauge
vertex.  Fig. 1g is the same as Fig. 1a but evaluated at zero
external momentum, contributing to $Z_{\mu}$. The evaluation of
these graphs is straightforward, since their singularity
structures involve only simple poles by power counting. Our
results for each graph agree with those in \cite{silva00}.
\renewcommand{\arraystretch}{1.5}
\begin{table}
\caption{Divergent contribution from each graph and to the appropriate renormalization constant.  Note that there are no divergences at one loop.
}
\begin{tabular}{llcllcllcll}
Graph&Divergence&\vline&Graph&Divergence&\vline&Graph&Divergence&\vline&Graph&Divergence\\
\hline
1a&$-\frac{g^4p^2}{6\pi^2\epsilon}$&\vline&1d&$-\frac{g^5}{24\pi^2\epsilon}$&\vline&1g&-$\frac{\mu g^4}{8\pi^2\epsilon}$&\vline&2c&$\frac{15wg^4}{8\pi^2\epsilon}$\\
1b&$-\frac{g^4p^2}{24\pi^2\epsilon}$&\vline&1e&$-\frac{g^5}{12\pi^2\epsilon}$&\vline&2a&$\frac{g^8}{\pi^2\epsilon}$&\vline&2d&$\frac{3wg^4}{2\pi^2\epsilon}$\\
1c&$-\frac{g^5}{6\pi^2\epsilon}$&\vline&1f&$\frac{g^5}{12\pi^2\epsilon}$&\vline&2b&$\frac{3g^8}{4\pi^2\epsilon}$&\vline&2e&$\frac{33w^2}{4\pi^2\epsilon}$\\
\end{tabular}
\end{table}

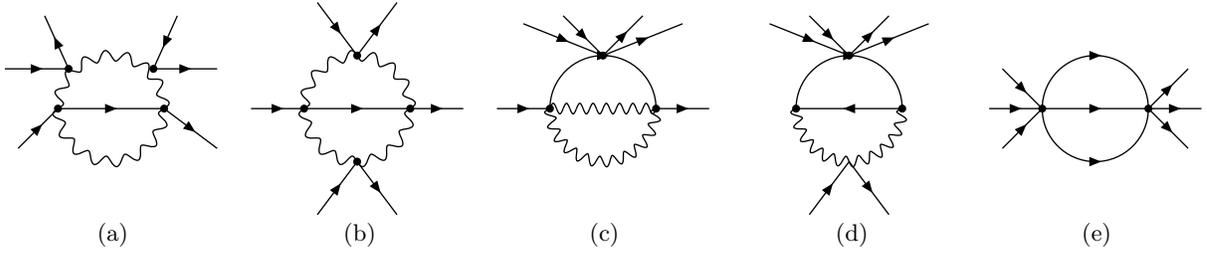
\begin{figure}
\begin{picture}(90,100)(0,0)
\ArrowLine(15,35)(30,50) \ArrowLine(30,50)(70,50)
\ArrowLine(70,50)(90,35) \PhotonArc(50,50)(20,0,180){2}{8}
\PhotonArc(50,50)(20,180,360){2}{8} \Vertex(30,50){1.5}
\Vertex(70,50){1.5} \ArrowLine(65.98,65)(90,65)
\ArrowLine(75,85)(65.98,65) \ArrowLine(10,65)(34.02,65)
\ArrowLine(34.02,65)(25,85) \Vertex(34.02,65){1.5}
\Vertex(65.98,65){1.5} \put(45,0){(a)}
\end{picture}
\begin{picture}(90,100)(0,0)
\ArrowLine(35,90)(50,70) \ArrowLine(50,70)(65,90)
\ArrowLine(35,10)(50,30) \ArrowLine(50,30)(65,10)
\PhotonArc(50,50)(20,0,180){2}{8}
\PhotonArc(50,50)(20,180,360){2}{8} \Vertex(50,30){1.5}
\Vertex(50,70){1.5} \ArrowLine(30,50)(70,50)
\ArrowLine(10,50)(30,50) \ArrowLine(70,50)(90,50)
\Vertex(30,50){1.5} \Vertex(70,50){1.5} \put(45,0){(b)}
\end{picture}
\begin{picture}(90,100)(0,0)
\Photon(30,50)(70,50){2}{8} \ArrowLine(10,50)(30,50)
\ArrowLine(70,50)(90,50) \PhotonArc(50,50)(20,180,0){2}{12}
\ArrowArcn(50,50)(20,180,360) \Vertex(30,50){1.5}
\Vertex(70,50){1.5} \Vertex(50,70){1.5} \ArrowLine(20,82)(50,70)
\ArrowLine(50,70)(80,82) \ArrowLine(35,85)(50,70)
\ArrowLine(50,70)(65,85) \put(45,0){(c)}
\end{picture}
\begin{picture}(90,100)(0,0)
\ArrowLine(70,50)(30,50) \ArrowLine(35,10)(50,30)
\ArrowLine(50,30)(65,10) \PhotonArc(50,50)(20,180,0){2}{12}
\ArrowArcn(50,50)(20,180,360) \Vertex(30,50){1.5}
\Vertex(70,50){1.5} \Vertex(50,70){1.5} \ArrowLine(20,82)(50,70)
\ArrowLine(50,70)(80,82) \ArrowLine(35,85)(50,70)
\ArrowLine(50,70)(65,85) \put(45,0){(d)}
\end{picture}
\begin{picture}(90,100)(0,0)
\ArrowLine(10,50)(30,50) \ArrowLine(30,50)(70,50)
\ArrowLine(70,50)(85,65) \ArrowLine(70,50)(85,35)
\ArrowLine(15,65)(30,50) \ArrowLine(15,35)(30,50)
\ArrowLine(70,50)(90,50) \ArrowArc(50,50)(-20,180,0)
\ArrowArc(50,50)(20,180,0) \Vertex(30,50){1.5} \Vertex(70,50){1.5}
\put(45,0){(e)}
\end{picture}
\caption{Two-loop contributions to $\Gamma^{(6,0)}$.  Figures (c),
(d) and (e) are representative of the topology; we have accounted
for other contractions of $\phi$ with $\phi^*$.}
\label{2Loop6Point}
\end{figure}

Employing the results in the table, we
have
\begin{eqnarray}
Z_\phi &=& 1 - \frac{5g^4}{24\pi^2\epsilon}\\
Z'_g &=& 1 - \frac{5g^4}{24\pi^2\epsilon}\\
Z_\mu &=& 1 - \frac{g^4 }{8\pi^2\epsilon}\\
Z_w&=& 1 + \frac{7 g^8}{4w\pi^2\epsilon} + \frac{27g^4 }{8\pi^2\epsilon}  +\frac{33w}{4\pi^2\epsilon}\\
Z_A&=& 1
\end{eqnarray}
from which we find that at two-loop order in $d=3-\epsilon$
\begin{eqnarray}
\beta_g&=&-\frac{\epsilon}{2}g\\
\eta_\phi &=& \frac{5g^4}{12\pi^2}\\
\beta_\mu&=& \mu\left(-2 + \frac{g^4}{6\pi^2}\right)\\
\beta_w&=&-w\left(2\epsilon - \frac{7g^8}{2w\pi^2} -
\frac{8g^4}{\pi^2} - \frac{33w}{2\pi^2}\right) =
\frac{33}{2\pi^2}(w-w_+)(w-w_-)
\end{eqnarray}
where
\begin{equation}
w_\pm(g)= \frac{1}{33}\left[2\pi^2\epsilon - 8g^4 \pm
\sqrt{4\pi^4\epsilon^2-32\pi^2g^4\epsilon-167g^8}\right]
\label{fpw}
\end{equation}
For nonvanishing $g$, as $\epsilon\rightarrow 0$ both roots are
complex and there is no physical $w$ fixed point.  However, for
$g^4\le0.851\epsilon$, the radicand in (\ref{fpw}) is real and
$w_+$ is a stable fixed point for $\epsilon>0$. Note, however,
that for $\epsilon>0$ the gauge coupling runs away to large values
and thus the only stable fixed point in $d=3-\epsilon$ is at
$(g,w)= (0,4\pi^2\epsilon/33)$. Focussing on $d=3$, we see that
$g$ is exactly marginal and $w_\pm = -g^4(0.242\mp 0.392i)$.
Writing $M=M_0e^{-\ell}$, we have
\begin{equation}
w(\ell) = \Re w_+ + \Im w_+\cot\left(\cot^{-1}\left[\frac{w_0 - \Re w_+}{\Im w_+}
\right]+\kappa\ell\right)
\end{equation}
where $\kappa = \frac{33}{2\pi^2 g^4}\Im w_+ = 0.655$.  Note that this solution is unstable and as
$\ell$ grows runs away to negative values of $w$, reminiscent of the behavior in a superconductor
\cite{HLM} and signaling a first-order transition. However, until any new fixed point controls the
scaling, the critical exponents $\nu$ and $\eta$ are determined entirely by the exactly marginal
coupling $g$:
\begin{eqnarray}
\eta&=&\frac{5g^4}{24\pi^2}\\
\nu&=&\frac{1}{2-g^4/6\pi^2}\approx\frac{1}{2}
+\frac{g^4}{24\pi^2}
\end{eqnarray}
We thus see that a chemical
potential for writhe can alter the radius of gyration exponent
$\nu$ and therefore writhe alters the universality class of a
self-avoiding walk before driving it to collapse.  The gauge field is not perturbatively
renormalized and thereby preserves its topological character.  As
in \cite{Kamien97} the scaling behavior of the average writhe
$\langle\Wr\rangle$ and the average squared writhe
$\langle\Wr^2\rangle$ are given in terms of the specific heat
exponent $\alpha=2-d\nu= \frac{1}{2} -\frac{g^4}{8\pi^2}$:
\begin{eqnarray}
\langle\,\Wr\,\rangle\sim -\frac{d}{d(g^2)} \ln\left(L^{\alpha-2}\right) &=&\frac{g^2}{4\pi^2}\ln L\\
\langle\,\Wr^2\,\rangle\sim
\frac{d^2}{d(g^2)^2}\ln\left(L^{\alpha-2}\right)&=&-\frac{1}{4\pi^2}\ln
L
\end{eqnarray}
we see that if the chemical potential vanishes then
$\langle\Wr\rangle=0$ as expected.  These logarithmic corrections
are in addition to the writhe that is stored is polymer segments
comparable to the persistence length ({\sl e.g.} plectonemes)
which scales as $L$, a result which does not violate the rigorous bound $\ln\langle\vert\Wr\vert\rangle\ge\frac{1}{2}\ln L$ \cite{Orlandini1994,Rensburg1993}.
Our expressions of scaling exponents $\eta$ and $\nu$ depend
continuously on $g^2$ and are thus similar to those in the
two-dimensional XY model.  There, topological defects are
responsible for this uncommon behavior;  here the topological link
constraint is responsible.

In conclusion, we have mapped the study of the statistics of
self-avoiding random walks in a $\Theta$ solvent to a
Chern-Simons field theory and have found a new scaling regime
for topologically constrained polymers by calculating the
scaling exponents $\eta$ and $\nu$ to two-loop order at the $\Theta$ point.
We conjecture that even in a good solvent the scaling behavior will be altered, though
as prior work indicates \cite{Kamien97} it is difficult to establish results in a controlled
approximation.  Since taking $N\rightarrow 0$ amounts to canceling a functional determinant, progress in this problem might be made by introducing fermionic partners to the complex scalars to have the
same effect.  It is possible that supersymmetric formulation of this field theory would yield a more complete understanding of the effect of topological constraints.

\begin{acknowledgments}
It is a pleasure to thank T.C. Lubensky and L. Radzihovsky for useful discussions.  This work was supported by NSF Grant DMR01-29804, the Donors of the Petroleum
Research Fund, Administered by the American Chemical
Society and a gift from L.J. Bernstein.
\end{acknowledgments}


\begin{thebibliography}{0}
\bibitem{Kamien97} J.D.~Moroz and R.D.~Kamien, Nucl. Phys. {\bf B
506}, 695 (1997).

\bibitem{White69}
J.H. White, Amer. J. Math. {\bf 91}, 693 (1969).
\bibitem{Fuller71}
F.B.~Fuller, Proc. Nat. Acad. Sci., USA, {\bf 68}, 815 (1971).

\bibitem{Fain97a}
B.~Fain and J~Rudnick, Phys. Rev. {\bf E 60}, 7239 (1999).

\bibitem{Fain97b}
B.~Fain, J.~Rudnick and S.~Ostlund, Phys. Rev. {\bf E 55}, 7364
(1997).

\bibitem{Julicher94}
F.~Julicher, Phys. Rev. {\bf E 49}, 2429 (1994).

\bibitem{Marko95}
J.F.~Marko and E.D.~Siggia, Phys. Rev. {\bf E 52}, 2912 (1995).

\bibitem{Gennes72}
P.G. de Gennes, Phys. Lett. {\bf A 38}, 609 (1972) 609.

\bibitem{Gennes75}
P.G. de Gennes, J. Phys. (France) Lett. {\bf 36}, 55 (1975) 55.

\bibitem{Duplantier82}
B.~Duplantier, J. Phys. (France) {\bf 43}, 991 (1982).

\bibitem{Duplantier87}
B.~Duplantier, J. Chem. Phys. {\bf 86}, 4233 (1987).

\bibitem{Grassberger}
P.~Grassberger, Phys. Rev. {\bf E 56}, 3682 (1997).

\bibitem{Young}
C.W.~Young, J.H.R.~Clarke, J.J.~Freire and m. Bishop, J. Chem.
Phys. {\bf 105}, 9666 (1996).

\bibitem{Rubio}
A.M.~Rubio and J.J.~Freire, J. Chem. Phys. {\bf 106}, 5638 (1997).

\bibitem{Hager99}
J.S.~Hager and L.~Schafer, Phys. Rev. {\bf E 60}, 2071 (1999).

\bibitem{Hager02}
J.S.~Hager, J. Phys. {\bf A 35}, 2703 (2002).

\bibitem{silva00}
L.C.~de~Albuquerque, M.~Gomes and A.J.~da~Silva, Phys. Rev. {\bf D
62}, 085005 (2000).

\bibitem{Chen92}
W.~Chen, G.W.~Semenoff and Y.S.~Wu, Phys. Rev. {\bf D 46}, 5521
(1992).

\bibitem{Semenoff88}
G.W.~Semenoff, P.~Sodano and Y.S.~Wu, Phys. Rev. Lett. {\bf 62},
715 (1988).

\bibitem{Niemi83}
A.J.~Niemi and G.W. Semenoff, Phys. Rev. Lett. {\bf 51}, 2077
(1983).

\bibitem{Redlich84}
A.N.~Redlich, Phys. Rev. {\bf D 29}, 2366 (1984).

\bibitem{Coleman85}
S.~Coleman and B.~Hill, Phys. Lett. {\bf B159}, 184 (1985).

\bibitem{HLM}
B.I. Halperin, T.C. Lubensky and S.-K. Ma, Phys. Rev. Lett. {\bf 32}, 292 (1974).

\bibitem{Orlandini1994}
E.~Orlandini, M.C.~Tesi, S.G.~Whittington, D.W.~Sumners and E.J.J.~van
Rensburg,
J. Phys. A: Math. Gen. {\bf 27}, L333 (1994).

\bibitem{Rensburg1993}
E.J.J.~van~Rensburg, E.~Orlandini, D.W.~Sumners, M.C.~Tesi and
S.G.~Whittington,
J. Phys. A: Math. Gen. {\bf 26}, L981 (1993).

\end{thebibliography}
\end{document}